# The Accuracy and Performance Analysis of the 1/t Wang-Landau Algorithm in the Joint Density of States Estimation

B. V. Kryzhanovsky[a,*], V. I. Egorov [a,**]

[a] *Scientific Research Institute for System Analysis of the RAS, Moscow*

*e-mail: kryzhanov@mail.ru*

**e-mail: rvladegorov@rambler.ru*

**Abstract**— The $1/t$ Wang-Landau algorithm is analyzed from the viewpoint of execution time and accuracy when it is used in computations of the density of states of a two-dimensional Ising model. We find that the simulation results have a systematic error, the magnitude of which decreases with increasing the lattice size. The relative error has two maxima: the first one is located near the energy of the ground state, and the second maximum corresponds to the value of the internal energy at the critical point. We demonstrate that it is impossible to estimate the execution time of the $1/t$ Wang-Landau algorithm in advance when simulating large lattices. The reason is that the criterion for switching to the $1/t$ mode was not met when the final value of the modification factor was reached. The simultaneous calculations of the density of states for energy and magnetization are shown to lead to higher accuracy in estimating statistical moments of internal energy.

## INTRODUCTION

The Wang-Landau (WL) algorithm [1] is an effective tool allowing direct evaluation of the density of states (DoS) of a system, i.e., the degeneracy degree of its energy states. The possibility of obtaining a joint DoS with respect to energy and other parameters, e.g., magnetization, is a particularly interesting characteristic of the algorithm. The use of this joint DoS makes it possible to get the probability density function of energy and magnetization at any temperature and external magnetic field. It is known that in the conventional realization of the algorithm, an error saturation occurs at a certain moment, and further calculations cannot increase the accuracy of DoS evaluation. In the case of the Ising model, it brings about a fluctuation error of the order of $10^{-4}$. Though the error seems small enough, the computation error for parameters measured in "real experiments", such as heat capacity, can be a considerable value. It is noteworthy that the paper [2] shows that during simulation, the modification factor should decrease by $1/t$ (where $t$ is the simulation time) to avoid the saturation of computation error. However, the analysis of the practicability of the $1/t$ approach in computations of the joint DoS has not been done so far.

Some authors (see [3]) point out that the major problem in WL algorithm-based calculations is that it is difficult to reach low-energy states in the process of the random walk. It results in a considerable increase in the algorithm execution time when simulating mid- and large-sized lattices. A possible way to overcome this problem is through parallelization of the algorithm and confining the random walk within a certain range of energy and magnetization. For instance, in papers [4, 5], the random walk involves manipulations with two spins of opposite signs, which helps keep the magnetization constant. However, the authors of papers [6, 7] point out that the division of the energy space leads to computation errors, which require configurations to be exchanged between sub-windows and overlapping of energy ranges.

In our research, we turn to examining the efficiency and accuracy of the $1/t$ WL algorithm in calculating conventional and joint DoSs for a two-dimensional Ising model. In that, we investigate the accuracy of computations of not only the density of states but also the heat capacity.

## CALCULATION OF THE ONE-DIMENSIONAL DENSITY OF STATES $g(E)$

To calculate the DoS $g(E)$ (i.e., the number of states with energy $E$) for an Ising $L \times L$ spin lattice with periodic boundary conditions, we use the algorithm described below. At first, we set the initial approximation of the DoS $g(E) = 1$, the initial modification factor $f = e$, and initialize the histogram $H(E) = 0$. Then we start a random walk across the system configurations by flipping a random spin. The probability of transfer to a new configuration $(i \rightarrow j)$ is

$$P(i \rightarrow j) = \min\left(\frac{g(E_i)}{g(E_j)}, 1\right) \qquad (1)$$

Independently of whether the new configuration is taken or not, the DoS and the histogram are updated for the energy of the current configuration:

$$g(E) \rightarrow f \cdot g(E), \qquad H(E) \rightarrow H(E) + 1. \qquad (2)$$

When the histogram is not zero $H(E) > 0$ for all energies, the modification factor decreases $f \rightarrow \sqrt{f}$ and the histogram is zeroed. The process continues until $\ln f > 1/t$, where $t = n/N_E$, $n$ is the total number of attempts to flip the spin from the start of the simulation, $N_E = L^2 - 1$ is the number of possible energy values. Once $\ln f \leq 1/t$, the simulation process changes: the modification factor updates at each simulation step according to the relationship $\ln f = 1/t$, and the state-visit histogram stops being used. The simulation ends when $\ln f < \ln f_{final} = 5 \cdot 10^{-8}$. The resulting DoSs are renormalized so that $g(E_0) = 2$ (where $E_0$ is the ground state). It should be noted here that the total number of iterations (attempts to flip a spin) is $N_E / \ln f_{final} = 2 \cdot 10^7 (L^2 - 1)$, which allows us to evaluate the algorithm execution time in advance.

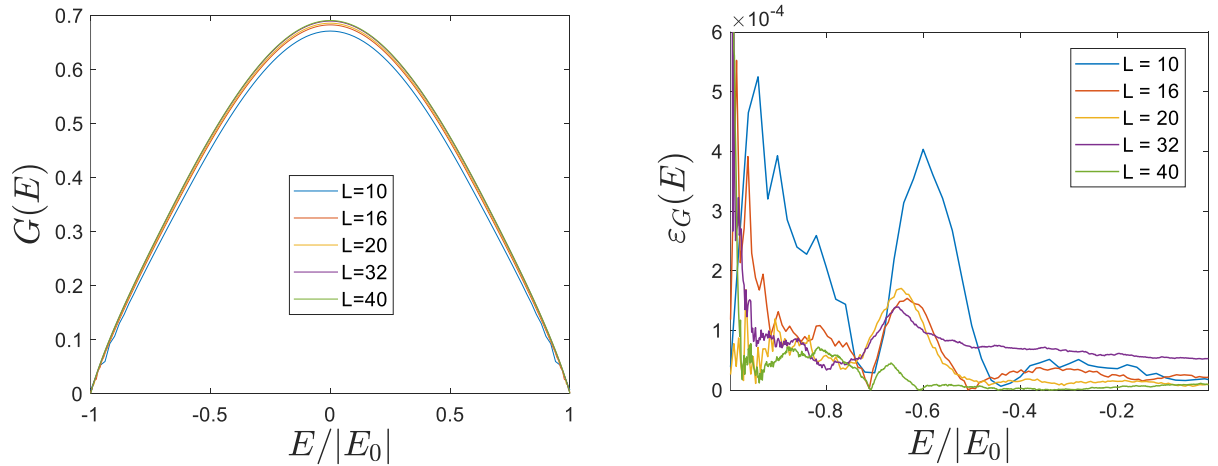

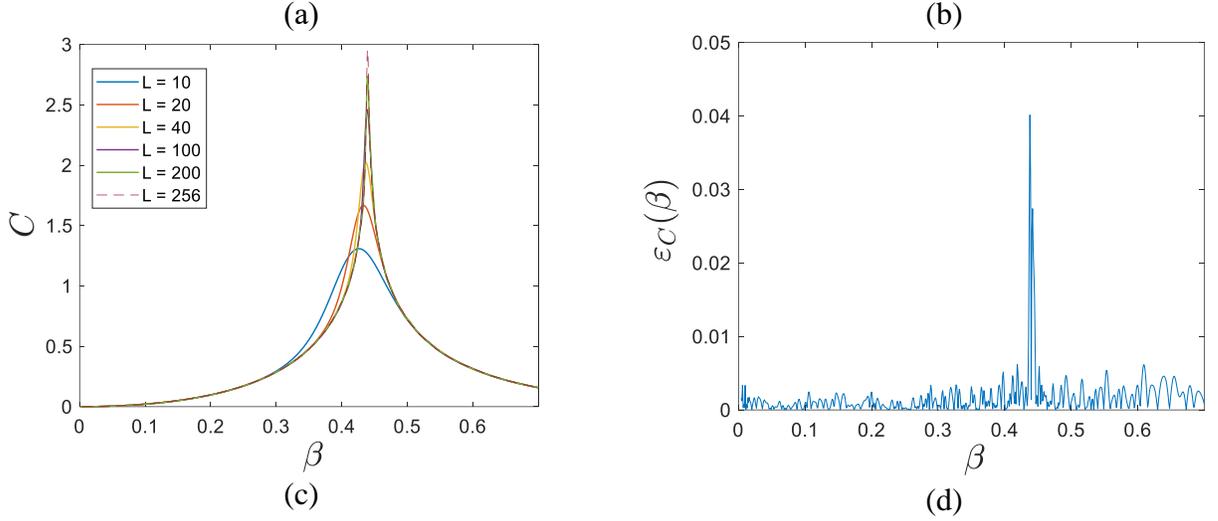

**Fig. 1.** One-dimensional DoSs $g(E)$. (a) the logarithm of DoSs $G(E)$ obtained by the simulation, (b) the relative error for $G(E)$ as compared to the exact value, (c) calculated temperature dependencies of heat capacity $C(\beta)$, (d) the relative error of the calculated heat capacity for $L = 256$.

We used this algorithm to calculate DoSs $g(E)$ for lattices with different $L$, from $L = 10$ to $L = 256$. Some of the dependencies $G(E) = \ln g(E)$ are given in Figure 1a. To evaluate the simulation accuracy, we compared the results with analytical dependencies for a two-dimensional Ising model determined by the method from the paper [8]. We also compare the temperature dependencies of heat capacity $C(\beta)$ ($\beta = 1/k_B T$ is the inverse temperature) obtained in the simulation (Fig. 1c) with exact dependencies derived from the Kauffman partition function [9]. Let us determine the relative error of the logarithm of the DoSs and heat capacity in the following way:

$$\varepsilon_G(E) = \left|\frac{G(E) - G_{ex}(E)}{G(E) + G_{ex}(E)}\right|, \quad \varepsilon_C(\beta) = \left|\frac{C(\beta) - C_{ex}(\beta)}{C(\beta) + C_{ex}(\beta)}\right|, \tag{3}$$

where $G_{ex}(E)$ and $C_{ex}(\beta)$ are exact values of logarithms of the DoSs and heat capacity.

The graphs of the relative error of the logarithm of the DoSs $\varepsilon_G(E)$ are shown in Figure 1b, and the mean values $\langle \varepsilon_G(E) \rangle$ and maximum values $\max(\varepsilon_G(E))$ of the relative error are presented in Table 1. As the method used in [8] does not allow the DoSs to be calculated for lattices with $L > 40$, it is impossible to evaluate the accuracy of the DoSs itself for large lattices. As is seen from Figure 1b, the error is largely a systematic one with relatively small fluctuations. For all dependencies, we can determine two points with the maximum error. The first point is near the ground state, the second point is near the critical value of the internal energy. Generally, the error decreases with the lattice size. The lattice with $L = 32$ is an exclusion. Yet in this case the unusually large error is the result of a poor normalization choice. Note that we tried to normalize the DoSs with respect to the total number of states, but this sort of normalization usually had worse agreement with exact values. The mean error $\langle \varepsilon_G(E) \rangle$ for most of the lattices is of the order of $10^{-5}$, and the maximum error $\max(\varepsilon_G(E))$ is of the order of $10^{-4}$ (see Table 1). At the same time, in general, $\langle \varepsilon_G(E) \rangle$ decreases with lattice size $L$, while $\max(\varepsilon_G(E))$ increases with $L$.

Also, we should note the specifics of the $1/t$ algorithm in computations of one-dimensional DoSs $g(E)$. Let us denote $t$ at which the simulation process changes (i.e., the passing from the conventional algorithm pattern to the $1/t$ algorithm) as $t_0$. It is seen from Table 1 that $t_0$ grows with lattice size $L$, i.e., passing to the $1/t$ algorithm occurs later. For a lattice with $L = 256$, the passing to the $1/t$ algorithm does not occur at all because condition $\ln f > 1/t$ keeps over the whole time of the simulation. That is why, for lattices with $L \geq 256$, it is impossible to evaluate the execution time of WL algorithm in advance given the specified finite value of the modification factor $f_{final} = \exp(5 \cdot 10^{-8})$.

**Table 1.** The results of the computation of one-dimensional DoSs $g(E)$. $L$ is the lattice size, $N_E$ is the number of possible energy values, $\langle \varepsilon_G(E) \rangle$ and $\max(\varepsilon_G(E))$ are the mean and maximum errors of the logarithm of the density of states, $\langle \varepsilon_C(\beta) \rangle$ and $\max(\varepsilon_C(\beta))$ are the mean and maximum errors of heat capacity. $t_0$ is the step of the simulation process at which the algorithm changes to the pattern $1/t$.

| $L$ | $N_E$ | $\langle \varepsilon_G(E) \rangle$ | $\max(\varepsilon_G(E))$ | $\langle \varepsilon_C(\beta) \rangle$ | $\max(\varepsilon_C(\beta))$ | $t_0$ |
|---|---|---|---|---|---|---|
| 10 | 99 | $1.4 \cdot 10^{-4}$ | $5.3 \cdot 10^{-4}$ | $5.2 \cdot 10^{-4}$ | $1.5 \cdot 10^{-3}$ | $7 \cdot 10^3$ |
| 12 | 143 | $9.7 \cdot 10^{-5}$ | $4.3 \cdot 10^{-4}$ | $3.4 \cdot 10^{-4}$ | $1.1 \cdot 10^{-3}$ | $6.5 \cdot 10^3$ |
| 14 | 195 | $8.2 \cdot 10^{-5}$ | $8.4 \cdot 10^{-4}$ | $5.5 \cdot 10^{-4}$ | $1.3 \cdot 10^{-3}$ | $1.5 \cdot 10^4$ |
| 16 | 255 | $6.7 \cdot 10^{-5}$ | $5.5 \cdot 10^{-4}$ | $3.7 \cdot 10^{-4}$ | $1.1 \cdot 10^{-3}$ | $1.4 \cdot 10^4$ |
| 18 | 323 | $3.2 \cdot 10^{-5}$ | $1.9 \cdot 10^{-4}$ | $5.4 \cdot 10^{-4}$ | $2.8 \cdot 10^{-3}$ | $2.3 \cdot 10^4$ |
| 20 | 399 | $4.4 \cdot 10^{-5}$ | $1.7 \cdot 10^{-4}$ | $4.1 \cdot 10^{-4}$ | $9.7 \cdot 10^{-4}$ | $2.6 \cdot 10^4$ |
| 32 | 1023 | $8.2 \cdot 10^{-5}$ | $1.2 \cdot 10^{-3}$ | $7.0 \cdot 10^{-4}$ | $3.3 \cdot 10^{-3}$ | $1.2 \cdot 10^5$ |
| 40 | 1599 | $3.3 \cdot 10^{-5}$ | $2.3 \cdot 10^{-3}$ | $6.7 \cdot 10^{-4}$ | $2.6 \cdot 10^{-3}$ | $1.2 \cdot 10^5$ |
| 100 | 9999 | - | - | $1.2 \cdot 10^{-3}$ | $7.5 \cdot 10^{-3}$ | $3.1 \cdot 10^6$ |
| 150 | 22499 | - | - | $2.4 \cdot 10^{-3}$ | $0.0146$ | $9.8 \cdot 10^6$ |
| 200 | 39999 | - | - | $1.9 \cdot 10^{-3}$ | $0.0569$ | $1.2 \cdot 10^7$ |
| 256 | 65535 | - | - | $1.7 \cdot 10^{-3}$ | $0.0401$ | - |

We see that the efficiency of the WL algorithm falls significantly with increasing lattice size $L$. Though the mean error of the DoS evaluation decreases, the error of the heat capacity calculated with the aid of this DoS grows. Moreover, a priori evaluation of the algorithm execution time becomes impossible for large lattices.

## CALCULATION OF THE JOINT DENSITY OF STATES $g(E, M)$

Now let us take use the $1/t$ WL algorithm to compute the joint DoS $g(E, M)$ with respect to energy $E$ and magnetization $M$. The simulation time is now defined as $t = n / N_{E,M}$, where $N_{E,M}$ is the total number of possible combinations of energy and magnetization values. $N_{E,M}$ (see Table 2) is determined by the results of a preliminary WL algorithm-based simulation without updating the modification factor. We could calculate DoS $g(E, M)$ only for small-sized lattices because $N_{E,M}$ is quite large and a great value of the simulation time $t$ is needed to attain the final modification factor. Figure 2a shows the graph $G(E, M) = \ln g(E, M)$ for the largest lattice ($L = 28$) that we considered. The normalization of $g(E, M)$ was also done with respect to the ground state.

**Table 2.** The results of computation of joint DoS $g(E, M)$. $N_{E,M}$ is the total number of possible combinations of energy and magnetization values. The other notation is the same as in Table 1.

| $L$ | $N_{E,M}$ | $\langle \varepsilon_G(E) \rangle$ | $\max(\varepsilon_G(E))$ | $\langle \varepsilon_C(\beta) \rangle$ | $\max(\varepsilon_C(\beta))$ | $t_0$ |
|---|---|---|---|---|---|---|
| 10 | 4200 | $7.8 \cdot 10^{-5}$ | $4.2 \cdot 10^{-4}$ | $8.8 \cdot 10^{-5}$ | $3.3 \cdot 10^{-4}$ | $7.8 \cdot 10^{3}$ |
| 12 | 8972 | $5.5 \cdot 10^{-5}$ | $3.0 \cdot 10^{-4}$ | $2.9 \cdot 10^{-5}$ | $1.1 \cdot 10^{-4}$ | $1.2 \cdot 10^{4}$ |
| 14 | 16982 | $5.2 \cdot 10^{-5}$ | $2.5 \cdot 10^{-4}$ | $8.0 \cdot 10^{-5}$ | $2.5 \cdot 10^{-4}$ | $2.4 \cdot 10^{4}$ |
| 16 | 29430 | $6.2 \cdot 10^{-5}$ | $5.8 \cdot 10^{-4}$ | $1.4 \cdot 10^{-4}$ | $6.2 \cdot 10^{-4}$ | $4.8 \cdot 10^{4}$ |
| 18 | 47724 | $4.3 \cdot 10^{-5}$ | $4.5 \cdot 10^{-4}$ | $1.5 \cdot 10^{-4}$ | $6.4 \cdot 10^{-4}$ | $5.1 \cdot 10^{4}$ |
| 20 | 73448 | $2.6 \cdot 10^{-5}$ | $2 \cdot 10^{-4}$ | $7.3 \cdot 10^{-5}$ | $3.1 \cdot 10^{-4}$ | $8.8 \cdot 10^{4}$ |
| 24 | 154530 | $3.0 \cdot 10^{-5}$ | $2.6 \cdot 10^{-4}$ | $5.2 \cdot 10^{-5}$ | $2.6 \cdot 10^{-4}$ | $1.6 \cdot 10^{5}$ |
| 28 | 289252 | $1.4 \cdot 10^{-5}$ | $1.3 \cdot 10^{-4}$ | $9.1 \cdot 10^{-5}$ | $5.7 \cdot 10^{-5}$ | $2.4 \cdot 10^{5}$ |

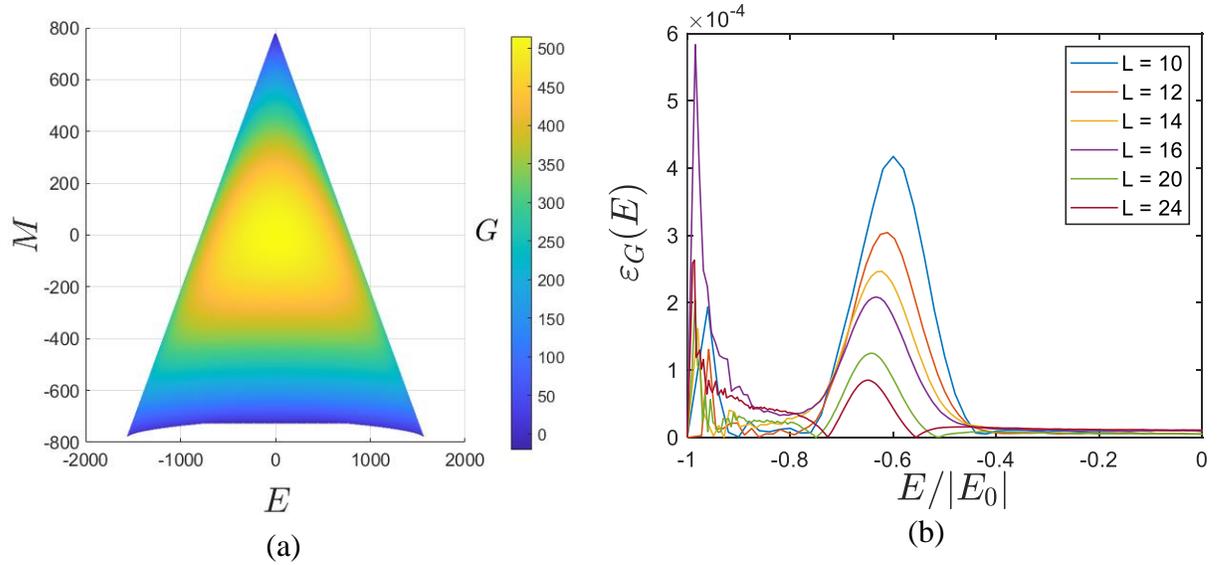

(a) (b)

**Fig. 2.** (a) The logarithm of the joint DoS $G(E, M)$. The data are simulation results of lattice $L = 28$. (b) The relative error of density $G(E)$ derived from $g(E, M)$ by summation as compared to the exact value.

Using joint DoS $g(E, M)$, we carried out the summation over $M$ and obtained the density of energy states $g(E) = \sum_M g(E, M)$. The graphs of the relative error of this density $\varepsilon_G(E)$ (see Figure

2b) are smoother than the results of direct calculations of the one-dimensional density. These dependencies have fluctuations only near the ground-state energy. The mean relative error $\langle \varepsilon_G(E) \rangle$ is almost half as much for most lattices comparing with previous results (see Table 2). Moreover, the evaluation accuracy for temperature dependencies of heat capacity is almost an order of magnitude as large (Table 2).

The exact values of joint density of states $g(E, M)$ for a two-dimensional Ising model are unknown, yet it is possible to determine mean energy $\bar{E}$, variance $D$ of energy and further statistical moments given a specified magnetization $M$ [11]. In the case of a two-dimensional Ising model the mean energy and energy variance are defined as

$$\bar{E}_{ex}(M) = E_0 \frac{(N-2n)^2 - N}{N(N-1)},$$

$$D_{ex}(M) = 2N \frac{(1-m^2)(1-m^2-4\delta+4\delta^2)}{(1-\delta)(1-2\delta)(1-3\delta)} \left(1 - \frac{4}{N-1}\right) \quad (4)$$

where $N = L^2$, $\delta = 1/N$, $m = M/N$, $n = (1-m)/2$.

Similarly to (3), for given $M$ we can determine the relative error of the mean energy and energy variance:

$$\varepsilon_{\bar{E}}(M) = \left|\frac{\bar{E}(M) - \bar{E}_{ex}(M)}{\bar{E}(M) + \bar{E}_{ex}(M)}\right|, \quad \varepsilon_D(M) = \left|\frac{D(M) - D_{ex}(M)}{D(M) + D_{ex}(M)}\right|, \quad (5)$$

where $\bar{E}(M)$ and $D(M)$ are the mean energy and energy variance derived from joint density $g(E, M)$. Mean errors $\langle \varepsilon_{\bar{E}}(M) \rangle$ and $\langle \varepsilon_D(M) \rangle$ for different lattices are given in Table 3. Clear that for all lattices the mean error is of the order of $10^{-4}$, i.e., the simulation results agree well with exact values.

**Table 3.** The mean energy error and energy variance given a fixed value of magnetization $M$.

| $L$ | $\langle \varepsilon_{\bar{E}}(M) \rangle$ | $\langle \varepsilon_D(M) \rangle$ |
|---|---|---|
| 10 | $6.4 \cdot 10^{-5}$ | $1.23 \cdot 10^{-4}$ |
| 12 | $1.7 \cdot 10^{-4}$ | $1.12 \cdot 10^{-4}$ |
| 14 | $1.1 \cdot 10^{-4}$ | $1.25 \cdot 10^{-4}$ |
| 16 | $8.4 \cdot 10^{-5}$ | $1.03 \cdot 10^{-4}$ |
| 18 | $1.2 \cdot 10^{-4}$ | $1.44 \cdot 10^{-4}$ |
| 20 | $1.1 \cdot 10^{-4}$ | $1.26 \cdot 10^{-4}$ |
| 24 | $1.3 \cdot 10^{-4}$ | $1.32 \cdot 10^{-4}$ |
| 28 | $1.8 \cdot 10^{-4}$ | $1.27 \cdot 10^{-4}$ |

In the conclusion of the paragraph let us note that $t_0$, i.e., the step at which the passing to pattern $1/t$ occurs, grows with the lattice size (Table 2). It means that in calculations of the joint DoSs for large lattices the evaluation of the algorithm execution time can become problematic, just as in computations of the one-dimensional density.

## CONCLUSION

We have examined the efficiency and accuracy of the WL algorithm when it follows a 1/t pattern and is used to calculate the one-dimensional DoS and joint DoS for a two-dimensional Ising model. The simultaneous calculations of DoS for energy and magnetization are shown to lead to higher accuracy of determination of statistical moments (e.g., variance). We found out that in computations of DoS Wang-Landau algorithm gives a systematic error which decreases with the lattice size. However, the efficiency of the algorithm falls considerably, and the evaluation of the algorithm execution time becomes impossible for most lattices.

The possible way to overcome these problems is to divide the system configuration space into domains. The most natural way is to carry out the simulation at a fixed value of magnetization by flipping two spins of opposite signs simultaneously. However, such division may be a cause for additional errors and, therefore, that would be a subject of our future study.

## FUNDING

The work was carried out within the framework of the state task of the Federal State Institution "Federal Scientific Center Scientific Research Institute for System Analysis of the Russian Academy of Sciences" on the topic FNEF-2024-0001 "Development and deployment of trusted AI systems based on new mathematical and algorithmic approaches and fast computing models compatible with domestic computer hardware" (1023032100070-3-1.2.1).

## CONFLICT OF INTEREST

The authors declare that they have no conflicts of interest.